\documentclass[12pt]{JHEP3}

\newcommand{\be}{\begin{equation}}
\newcommand{\ee}{\end{equation}}
\newcommand{\bea}{\begin{eqnarray}}
\newcommand{\eea}{\end{eqnarray}}

\def\nn{\nonumber}
\def\la{\langle}
\def\ra{\rangle}
\def\de{\partial}

\def\sla{\raise.15ex\hbox{$/$}\kern-.57em}
\def\ie{{\it i.e.}~}
\def\i{{\rm i}}
\def\eg{{\it e.g.}~}
\def\ap{{\alpha^\prime}}

\def\a{\alpha}
\def\b{\beta}

\def\m{\mu}
\def\n{\nu}

\def\cA{{\cal A}}
\def\cB{{\cal B}}
\def\cC{{\cal C}}

\def\cE{{\cal E}}
\def\cF{{\cal F}}

\def\cH{{\cal H}}
\def\cI{{\cal I}}
\def\cJ{{\cal J}}
\def\cK{{\cal K}}

\def\cM{{\cal M}}
\def\cN{{\cal N}}

\def\cP{{\cal P}}
\def\cQ{{\cal Q}}
\def\cR{{\cal R}}

\def\cT{{\cal T}}
\def\cU{{\cal U}}

\def\cW{{\cal W}}
\def\cX{{\cal X}}

\author{Pascal Anastasopoulos, Massimo Bianchi, Gor Sarkissian and Yassen S. Stanev \\
Dipartimento di Fisica \& Sezione I.N.F.N.\\
Universit\`a di Roma
``Tor Vergata'' \\
Via della Ricerca Scientifica, 1 - 00133  Roma ITALY\\
E-mail: \email{Pascal.Anastasopoulos@roma2.infn.it,}\\
~~~~~~~~~ \email{Massimo.Bianchi@roma2.infn.it,}\\
~~~~~~~~~ \email{Gor.Sarkissian@roma2.infn.it,}\\
~~~~~~~~~ \email{Yassen.Stanev@roma2.infn.it}.}
\abstract{We derive general formulae for tree level gauge
couplings and their one-loop thresholds in Type I models based on
genuinely interacting internal $\cN =2$ SCFT's, such as Gepner
models. We illustrate our procedure in the simple yet non-trivial
instance of the Quintic. We
briefly address the phenomenologically more relevant issue of
determining the Weinberg angle in this class of models. Finally we
initiate the study of the correspondence between `magnetized' or
`coisotropic' D-branes in Gepner models and twisted
representations of the underlying $\cN=2$ SCA.}

\preprint{ROM2F/2006/29}

\title{On gauge couplings and thresholds in Type I Gepner models and otherwise}

\begin{document}

\section{Introduction}

Type I strings and their close relatives
have received a great deal of attention in the past few years (see \eg
\cite{Dudas:2000bn}-\cite{Blumenewreview} for
comprehensive reviews).

Although their systematization was already achieved in the early
90's \cite{Sagnotti:1987tw}-\cite{Bianchi:1997rf}, including the
possibilities of minimally coupling R-R p-form potentials and
reducing the rank of the Chan-Paton group by turning on a
quantized NS-NS antisymmetric tensor background
\cite{Bianchi:1991eu,Bianchi:1997rf}, the geometric description in
terms of D-branes and $\Omega$-planes \cite{Polchinski:1995mt,
Gimon:1996rq}, pioneered in \cite{Dai:1989ua, Leigh:1989jq} has
definitely consecrated this framework as the most promising one to
embed Particle Physics in String Theory. Simple instances of
chiral model based on toroidal orbifolds
\cite{Berkooz:1996dw}-\cite{Bianchi:1999uq} with or without
intersecting branes
\cite{Berkooz:1996km}-\cite{Angelantonj:2005hs}, that are T-dual
to magnetized branes
\cite{Balasubramanian:1996uc}-\cite{Dudas:2005jx}, represent a
useful guidance for more sophisticated and hopefully realistic
constructions that may require {\it inter alia} (non) commuting
open string Wilson lines or their closed string dual constructions
\cite{Bianchi:1997rf}-\cite{Cvetic:2001nr}.

The important issues of supersymmetry breaking
\cite{Bachas:1995ik}-\cite{Angelantonj:2004yt} and moduli
stabilization \cite{Giddings:2001yu}-\cite{Camara:2005dc} have
been tackled with some success.  Interactions at tree (disk and
sphere) level \cite{Klebanov:2003my}-\cite{Abel:2004ue} have been
studied in some detail. One-loop thresholds for the gauge
couplings have been computed
\cite{Bachas:1992bh}-\cite{Bianchi:2005sa} and some steps beyond
one-loop have been made \cite{Berkovits:2006bk}. More recently it
has been argued that large extra dimensions naturally emerge in
this approach \cite{Antoniadis:1998ig}-\cite{Dienes:1998vg}. In
these cases, predictions for processes with missing energy at near
future colliders \cite{Dudas:1999gz}-\cite{Chialva:2005gt} have
been put forward.

Following the by now standard construction of RCFT's on surfaces
with crosscaps and boundaries
\cite{Pradisi:1995qy}-\cite{Pradisi:1996yd}, open and unoriented
models based on genuinely interacting internal $\cN =2$ SCFT's,
such as Gepner models \cite{Gepner:1987vz}-\cite{Eguchi:1988vr}, have
been constructed in
\cite{Angelantonj:1996mw}-\cite{Dijkstra:2004cc} and accurately
scanned  in order to test the possibility of accommodating the
Standard Model \cite{Dijkstra:2004ym, Anastasopoulos:2006da}.
Indeed, contrary to perturbative heterotic strings, it is rather
contrived if not impossible to embed interesting Grand Unified
Theories (GUT's) in perturbative Type I strings. Exceptional groups, such
as $E(6)$ are ruled out by a theorem of Marcus and Sagnotti's
\cite{Marcus:1982fr, Marcus:1986cm}, and the same applies to
spinorial representation of Orthogonal groups, such as $SO(10)$.
One could then look for chiral GUT's based on unitary groups such as
$SU(5)$. Although, with some effort, one can find reasonable
$U(5)$ three generation models with Higgses in the adjoint and in
the $5 + \bar{5}$, these models turn out to be unrealistic since
only the Yukawa couplings $\phi_{\bf 5} \psi_{\bf 5} \chi_{\bf
\bar{10}}$ are allowed by $U(1)$ charge conservation. The Yukawa
couplings $\phi_{\bf \bar{5}} \psi_{\bf \bar{10}} \chi_{\bf
\bar{10}}$, though $SU(5)$ invariant, are forbidden by $U(1)$
charge conservation and by the impossibility of generating the
necessary antisymmetric tensor $\epsilon_{ijklm}$ as Chan-Paton
factor, \ie taking traces of matrices  \cite{Antebi:2005hr,Berenstein:2006pk}. Barring non-perturbative
effects that can significantly change this state of affairs but
whose study is only in its infancy, the best one can achieve is
some L-R symmetric extension of the SM or a Pati-Salam
generalization thereof, together with some (anomalous) $U(1)$'s.
The role of the latter has been carefully studied recently
\cite{Anastasopoulos:2006cz,Anastasopoulos:2007qm} and we will not add much here.

Aim of the present paper is to derive general formulae for the
(non abelian) gauge couplings and their one-loop thresholds in Type I models
based on type II Gepner models. Quite remarkably we will find
elegant and compact formulae valid whenever the internal CFT
enjoys $\cN =2$ worldsheet SCI. This allows to construct a parent
type II (B) theory which is supersymmetric and corresponds to
the compactification on a CY 3-fold (or $K3$ or $T^2$). Depending on
the brane and $\Omega$ plane configuration, the resulting type I
model may enjoy spacetime susy. Indeed, as it was first advocated
in \cite{Bianchi:2000vb} and it was exploited more recently in the
context of Black-Hole physics \cite{Emparan:2006it} and
intersecting D-brane models \cite{Floratos:2006hs}, it is possible
that each pair of branes enjoys some susy (common to the
$\Omega$-planes that can in fact coincide with some of the stacks)
which is not the same for all pairs. Even in this case, one-loop
amplitudes would look supersymmetric and some of the threshold
corrections could be reliably computed by means of our formulae.

After illustrating our formulae in the case of a Type I model on
the Quintic with gauge group $SO(12)\times SO(20)$, we
address the possibility of determining the Weinberg angle in
phenomenologically more promising models in this
class. This is tightly related to the embedding of the
$U(1)_Y$ hypercharge generator in the Chan-Paton group
\cite{Anastasopoulos:2006da}.

Finally, we briefly discuss the issue of computing some four-point
amplitudes along the lines of \cite{Bianchi:2006nf} and initiate
the program of studying and classifying `magnetized' or
`coisotropic' D-branes in Gepner models. As it was argued in
\cite{Bianchi:2005yz}, these correspond to twisted representations
of the underlying $\cN=2$ superconformal algebra (SCA). We will
not explicitly consider the interesting possibility of
constructing models with large extra dimensions based on (freely
acting) orbifolds of $K3 \times T^2$ at Gepner points for $K3$
\cite{Aldazabal:2003ub}. Neither we will consider turning on
closed string fluxes (metric torsion, NS-NS 3-form flux and R-R
fluxes) and their effect of  non-trivial warping of the geometry
\cite{Grana:2005jc}. Being optimistic, this would at least require
resorting to alternative approaches \cite{Linch:2006ig}, where
supersymmetry properties are manifest such as the pure spinor
formalism \cite{Berkovits:2001nv}, or the hybrid formalism
\cite{Berkovits:1996bf} or other manifestly supersymmetric
formalisms \cite{Berkovits:2006bk, Mafra:2005jh}.

We leave to future work a more thorough analysis of gauge
couplings and thresholds in phenomenologically viable models as
well as the study of other important ingredients in the low-energy
effective action.

{}{}{}{}{}{}{}{}{}{}

\section{$\cN =2$ SCFT and Gepner models}

We start with a general discussion of the worldsheet properties of
supersymmetric vacuum configurations for (open and unoriented)
strings.

\subsection{$\cN =2$ SCFT}

As it was shown by Banks and Dixon
\cite{Banks:1988yz} in order to have spacetime susy
in $D=4$, the underlying SCFT must enjoy at least $\cN =2$
superconformal invariance on the worldsheet. In addition to the
stress tensor $T$ and the two spin 3/2 supercurrents $G^+$ and
$G^-$, the $\cN =2$ superconformal algebra includes also a $U(1)$
R-symmetry current $J$. {\it A priori} the $\cN=2$ worldsheet
supercurrents can acquire arbitrary phases under parallel
transport around non-trivial cycles \ie
\be G^\pm
(e^{2\pi {\rm i}} z) =  e^{2\pi {\rm i} \nu_{\pm}} G^\pm (z) \, \
\ee
where we can choose $|\nu_{\pm}| \le 1/2$. As a consequence, their
modes are labelled by $ r_{\pm} \in {\bf Z} +1/2 + \nu_{\pm}$. If
$\nu_{+} + \nu_{-} \neq 0$ the current $J$ has non-integer modes
and one finds what is called a `twisted' representation of the
$\cN=2$ SCA. In the rest of the paper we shall consider only the
case $\nu_{\pm} =\pm \nu$. As a consequence, the current $J$ has
integer modes and the two supercurrents $G^{\pm}$ have $U(1)$
charge $\pm 1$ respectively. Different values of $\nu$ are
isomorphic and they are connected by the `spectral flow' induced
by the action of the unitary operator
\be \cU_{\nu} = \exp(2\pi \i \nu J_0 ) \ . \ee
The cases $\nu = 0$ and $\nu=\pm 1/2$ correspond to the NS and R
sector, respectively, and they are related by one unit of spectral
flow \ie by $ \cU_{\pm 1/2}$. These are singled out as the only
boundary conditions compatible with the `reality' of the $\cN=1$
supercurrent
\be G_{\cN =1}  = G^+ + G^- \ , \ee
that couples to the worldsheet gravitino. Bosonizing the $U(1)$
current as
\be J ={\rm i} {\sqrt{c \over 3}} \de H  \ , \ee
the spectral flow is related to the spacetime supercharges
\cite{FriedanMartinecShenker86} that in $D=4$ read~\footnote{We mostly focus
on the case $D=4$ corresponding to $c_{int}=9$.}
\be \cQ_\a = \int {dz \over 2\pi \i} e^{-\varphi/2} S_\a
e^{{\i\over 2} \sqrt{c \over 3} H } \quad , \quad \cQ_{\dot\a} =
\int {dz \over 2\pi \i} e^{-\varphi/2} C_{\dot\a} e^{-{\i\over 2}
\sqrt{c \over 3} H} \, \ee
where $\varphi$ is the superghost boson and $S_\a, C_{\dot\a}$ are
spin fields of opposite chirality. Depending on the spin of the
state, locality of the OPE of $\cQ_\a, \cQ_{\dot\a}$ with the
vertex operators determines the correct quantization condition for
the $U(1)$ R-charge.

In $D=4$, \ie for $c_{int}=9$, the vertex operator for a vector
boson is
\be V_{-1} =
a_\mu(p) \psi^\mu e^{-\varphi} e^{{\i q_v} \sqrt{3 \over c} H }
e^{\i pX} \, \ee
locality requires $q_v = 0 \, ({\rm mod} \, 2)$. The vertex
operator for a scalar is
\be V_{-1} = \phi(p) e^{-\varphi} \widehat\Psi_{q_o} e^{\i
q_o \sqrt{3 \over c} H } e^{\i pX} \ , \ee
where $\Psi_{q_o} = \widehat\Psi_{q_o} e^{\i q_o \sqrt{3 \over c}
H }$ is a primary field in the NS sector with $U(1)$ charge
$q=q_o$ and dimension $h = (1 + p^2)/2$. Locality requires $q_o =
1 \, ({\rm mod} \, 2)$. Massless scalars correspond to
(anti)chiral primaries with $h=1/2$ so that $q_o = \pm 1$, {\it a
priori} $0 \le q_{_{CPO}} <c/3$. For the (massless) LH spinor, the
vertex operator is
\be V_{-1/2} = u^\a(p) S_\a e^{-\varphi/2}
\widehat\Sigma_{q_s} e^{\i q_s \sqrt{3 \over c} H } e^{\i pX} ,
\ee
where $\Sigma_{q_s} = \widehat\Sigma_{q_s} e^{\i q_s \sqrt{c \over
3} H }$ is a primary field in the R sector with $U(1)$ charge
$q=q_s$ and dimension $h = (c/24) + (p^2/2)$ and locality requires
$q_s = +3/2 \, ({\rm mod} \, 2)$. Massless LH spinors correspond
to R groundstates (RGS) with $h = c/24 = 3/8$ and $q_s = +3/2$,
so that $h_{\widehat\Sigma} = (c/24) - (3q_s^2/2c)$, {\it a
priori} $-c/6 \le q_{_{RGS}} \le c/6$. For the (massless) RH
spinor, the vertex operator is
\be V_{-1/2} = v_{\dot\a}(p) e^{-\varphi/2}
C^{\dot\a} \widehat\Sigma^\dagger_{q_c} e^{\i q_c \sqrt{3 \over c}
H } e^{\i pX} \ee
where locality requires $q_c = -3/2 \, ({\rm mod} \, 2)$.

{}{}{}{}{}{}{}{}{}{}

\subsection{Unitary $\cN =2$ minimal models}

Unitary $\cN =2$ minimal models are known to form a discrete series
\cite{FriedanMartinecShenker86}.
They are equivalent to the quotients $SU(2)_k \times U(1)_2/
U(1)_{k+2}$, so that the central charge is given by
\be c(k) = {3k \over k+2} \, , \ee
where $k$ is a positive integer. The $\cN =2$ primary fields
$\Phi^{\ell (k)}_{m,s}$ are labelled by three quantum numbers
$0\le \ell \le k$, $ -(k+1) \le m \le k+2$ and $s=0,\pm 1, 2$,
with $\ell + m + s = 0 \, ({\rm mod} \,  2)$. By the field
identifications
\be
\Phi^{\ell (k)}_{m,s} = \Phi^{\ell (k)}_{m,s+4} = \Phi^{\ell
(k)}_{m + 2(k+2),s} = \Phi^{k - \ell (k)}_{m+ k+2,s+2} \ee
one can restrict the values of $(\ell, m, s)$ to the `standard'
range $s = 0, \pm 1$, $\ell \le [k/2]$, $ -(k+1) < m \le k+2$.

The spectrum of conformal dimensions and $U(1)$ charges are given
by~\footnote{In the rest of this paper we shall always tacitly
assume the $({\rm mod}\ 1 )$ and $({\rm mod}\ 2 )$ conditions for
$h$ and $q$.}
\bea &&h(\ell, m, s)
= {\ell(\ell + 2) - m^2 \over 4(k+2)} + {s^2\over 8} \ \ ({\rm mod}
\ 1 )\ ,\\
&&q (m,s) = {m \over k+2} - {s \over 2} \ \ ({\rm mod} \ 2 ) \
. \eea

Every $\cN =2$ minimal model can be decomposed into a
parafermionic theory and a free $U(1)$ boson, so that
\be T = T_{PF} - {1\over 2} \de H \de H \quad , \quad G^\pm =
\sqrt{2 c(k)\over k} \psi^\pm_{PF} e^{\pm \i\sqrt{3\over c(k)}H}
\quad , \quad J ={\rm i} {\sqrt{c(k) \over 3}} \de H \ee
one has
\be \Phi^{\ell(k)}_{m,s} =
\hat\Phi^{\ell(k)}_{m-s} e^{\i \gamma^k_{m,s} H} \ee
where
\be
\gamma^k_{m,s} = \sqrt{k+2 \over k} \left( {m\over k+2} -  {s
\over 2}\right) = \sqrt{3\over c(k)} \, q(m,s) \, . \ee
Unitarity requires $ h_{\hat\Phi}\ge 0$ \ie $h_{\Phi} \ge
3q^2/2c(k)$ (`unitary parabola'). Moreover, in the NS sector
($s=0,2$)
\be h_{_{NS}} \ge
{1\over 2} |q_{_{NS}}| \ . \ee
The inequality is saturated by (anti) chiral primary operators
(CPO) corresponding to $m = \pm \ell$ and $s=0$ with
$|q_{_{CPO}}|\le c(k)/3$ that satisfy
\be G^+_{-
1/2}|h = q/2;  \  q\ra_{_{CPO}} = 0 \quad {\rm or} \quad G^-_{-
1/2}|h = - q/2; \  q\ra_{_{{CPO}^\dagger}} = 0 \ . \ee
In the R
sector ($s=\pm 1$)
\be h_{R} \ge {c(k) \over 24} \ . \ee
The inequality is saturated by Ramond ground-states (RGS)
corresponding to $m = \pm (\ell+1)$ and $s=\pm 1$ with
$|q_{_{RGS}}|\le c(k)/6$ that satisfy
\be G^\pm_{0}|h = c(k)/24; \
q\ra_{_{RGS}} = 0  \ee
and contribute to the Witten index $\cI_W = Tr(-)^F$.

{}{}{}{}{}{}{}{}{}{}

\subsection{Gepner models}

Gepner models \cite{Gepner:1987vz, Eguchi:1988vr} are tensor
products of $r$ minimal $\cN=2$ models quotiented by a subgroup of
the discrete symmetries that keeps only the states with quantized
$U(1)$ charge and sectors in which the $\cN=1$ worldsheet
supercurrent
\be G_{\cN =1}  = G^+ + G^- = \sum_{i=1}^r (G^+_i + G^-_i)
\label{wssusydec}\ee
is well defined \ie transforms covariantly, acquiring at most a
sign under parallel transport around non-trivial cycles. The
latter condition looks at first as a merely worldsheet
requirement, dictated by consistency of the coupling of $G_{\cN
=1}$ to the worldsheet gravitino, but actually it is a necessary
condition for BRS invariance and decoupling of negative norm
states. The $U(1)$ charge quantization is equivalent to the
condition for spacetime supersymmetry, whose chiral action
(`spectral flow') is only well defined on states with quantized
$U(1)$ charges. Indeed bosonizing the $U(1)$ current one finds
\be \cJ = {\rm i}  {\sqrt{c \over 3}} \de H = {\rm i} \sum_i
{\sqrt{c_i \over 3}} \de H_i \ , \ee so that \be H = {\rm i}
\sum_i {\sqrt{c_i \over c}} \de H_i \ . \ee

For our latter purposes it is crucial to further investigate the
decomposition of the individual terms in the $\cN=1$ worldsheet
supercurrent (\ref{wssusydec})
\be G_i^\pm = \widehat{G}_i^\pm e^{\pm \i \sqrt{3 \over c_i} H_i }
\label{wssupcurri}\ee
where $\widehat{G}_i^\pm = \psi^\pm_{PF,i}$ are NS primary fields
of dimension
\be h_{\widehat{G}_i^\pm} = {3\over 2}- {3\over 2c_i} = 1- {1\over
k_i} \ee
that can be identified with the fundamental ${\bf Z}_{k_i}$
parafermions defining the coset $SU(2)_{k_i}/U(1)$. In particular
for $k = 1$ one has $\widehat{G}^\pm  = {\bf 1}$, while for $k =
2$ one finds $\widehat{G}^+ = \widehat{G}^-  = \psi$, the `real'
fermion of the Ising model. The first `non-trivial' case is $k =
3$ (relevant for the quintic) where $\widehat{G}^+  = \rho$ and
$\widehat{G}^- = \rho^\dagger$ with $h_{\rho} = h_{\rho^\dagger} =
2/3$.

Moreover, as stated above, one should demand  $\nu_i = \nu_j =
\nu_{st}$ for any $i$ and $j$, with $st$ standing for $G_{st} =
\psi_\mu \de X^\mu$, in order for $G_{tot} = G_{st} + G_{int}$ to
be well defined. This is at the heart of the so-called $\beta_i$
${\bf Z}_2^r$-projections which were first introduced by Gepner
\cite{Gepner:1987vz} in analogy with what was done in free
fermionic models. Experience with orbifolds and
magnetized and/or intersecting D-branes suggest that `twisted'
representations that have been thrown out of the door may snick in
through the window. Indeed, one can preserve covariance of the
$\cN=1$ worldsheet supercurrent $G_{tot} = G_{st} + G_{int}$ by
changing the boundary conditions of the internal bosonic
coordinates $X^I_{int}$ and at the same time by implementing the
same (`contragradient') change of the internal fermionic
coordinates $\Psi^I_{int}$. Inspection of (\ref{wssupcurri})
suggests that a shift of the boson $H_i$ can be `compensated' by a
twist of the parafermion $\psi_{PF,i}$. We will further elaborate
on this observation in Section \ref{magncoiso}. For the time being
let us focus on standard `untwisted' UIR's.

For tensor product theories, primary fields can be written as
$\Phi_{h,q} = \prod_i \Phi_{h_i,q_i}$ with
\be h = \sum_{i=1}^r h_i = \sum_{i=1}^r \left[ {\ell_i(\ell_i + 2)
- m_i^2 \over 4(k_i+2)} + {s_i^2\over 8} \right] \ee
and
\be q = \sum_{i=1}^r q_i = \sum_{i=1}^r \left[ {m_i \over k_i+2} -
{s_i\over 2} \right] \ . \ee
In order to restrict the spectrum to the states on which $G_{\cN
=1}$ acts consistently, as a whole, one has to combine states that
impose the same boundary condition on each term in
(\ref{wssupcurri}). The resulting ${\bf Z}_2^r$ projection can be
achieved in different ways. We follow the orbit procedure
developed by Eguchi, Ooguri, Taormina and Yang
\cite{Eguchi:1988vr}.

In our conventions, the total susy charge (which is proportional to the
spectral flow operator) reads
\be Q = S_2 \prod_i\Phi^{(i), 0}_{-1, -1} \, . \ee
It has total charge $c / 6 = \sum_i c_i / 6 = 3/2$ in $D=4$ \ie
for $c=9$ and, barring the spin field $S_2$ (with helicity $
\lambda = +1/2$ and scaling dimension 1/8), dimension ${c / 24} =
\sum_i {c_i / 24} = {3 / 8}$. The supercurrent in each subtheory
reads
\be G_i = \Phi^{(i), 0}_{0, 2} \, . \ee
Then, given the Highest Weight State (HWS) $\cX^{HWS}_V$ in the
$V_2$ part of a `resolved' susy character (\eg for the identity
sector $\cX^{HWS}_V = \prod_i \cX^{(i), 0}_{0, 0}$) satisfying
\be q^{HWS}_{V} = 0 \ ({\rm mod } \ 2) \label{qhws} \ee
the action of $Q^n \prod_i G_i^{p_i}$ maps it into a state with
\be q(n, p_i) = q^{HWS}_{V} + n {c \over 6} + \sum_i p_i =
            q^{HWS}_{V} + n {3 \over 2} + \sum_i |p_i| \, ,
\ee
since $p_i = |p_i|$ (mod 2) for $p_i = 0,1$. Setting
\be K = {\rm l.c.m.}\{4, 2(k_i + 2)\} \ee
we can write for the complete Gepner model characters
(orbits)\footnote{When all the levels are odd, one has in addition
to divide (\ref{chiI}) by 2.}
\be \chi_I = \sum_{n=0}^{K-1} (-)^n  Q^n \ \prod_{i=1}^{r} \ \
\sum_{p_i=0,1} (O_2 \, G_i)^{p_i} ( V_2 \, \cX_{V,I}^{HWS}) \ .
\label{chiI}\ee

For some purposes, it is convenient to manifestly separate the
contribution of the non compact space-time super-coordinates and
write \eg for $D=4$
\be
 \chi_I =  V_2 \cX^V_I  + O_2 \cX^O_I - S_2 \cX^S_I  - C_2 \cX^C_I \, ,
\label{susyChi}  \ee
where $V_2$, $O_2$, $S_2$, $C_2$ are $SO(2)$ characters at level
one and the minus signs take into account spin and statistics.
Supersymmetry entails $\chi_I = 0$ for all $I$. Since by
assumption (see eq.~(\ref{qhws}))
\be q^{HWS}_V = \sum_i \left[ {m_i \over k_i + 2} - {s_i \over 2}
\right] = 0 \ ({\rm mod}  \, 2) \ee
for the Ramond sector internal characters one finds
\be \cX^S_I = \sum_{m=0}^{{K \over 4} -1} \left[\sum_{p_i=0,1
\atop \sum_i p_i = 0 (mod \,2) }\prod_i \chi^{\ell_i (k_i)}_{m_i -
4 m - 1, s_i - 1 + 2p_i} + \sum_{p_i=0,1 \atop \sum_i p_i = 1 (mod
\, 2) }\prod_i \chi^{\ell_i (k_i)}_{m_i - 4 m + 1, s_i + 1 + 2p_i}
\right] \, , \label{ChiS} \ee
and
\be \cX^C_I = \sum_{m=0}^{{K \over 4} -1} \left[ \sum_{p_i=0,1
\atop \sum_i p_i = 0 (mod \, 2) }\prod_i \chi^{\ell_i (k_i)}_{m_i
- 4 m + 1, s_i + 1 + 2p_i} + \sum_{p_i=0,1 \atop \sum_i p_i = 1
(mod \, 2) }\prod_i \chi^{\ell_i (k_i)}_{m_i - 4 m - 1, s_i - 1 +
2p_i} \right] \, . \label{ChiC} \ee

In principle the resulting `supersymmetric' characters
(\ref{susyChi}) have `length' (total number of terms) $L = 2^r K$.
However, although neither $Q$ nor $G_i$ independently have fixed
points, \ie they act freely, it may happen that some orbits are
shorter due to field identifications. It can be shown, that this
can happen only when some of the $k_i$ are even and that the short
orbits are always twice shorter $L_{short} = 2^{r-1} K$, so as far
as only the spectrum of conformal dimensions is considered, one
has simply to halve the expressions above. The situation becomes
more involved if modular transformations are considered. In this
case one has to resolve the fixed point ambiguity which amounts to
`split' the representation encoded in the supersymmetric character
into two independent representations, possibly conjugate to one
another, that have to be labelled by an additional quantum number.

{}{}{}{}{}{}{}{}{}{}

\subsection{Open descendants}

In these cases, the parent `oriented' closed string theory is
based on a perturbative spectrum encoded in the one-loop torus
partition function
\be {\cal T} = \sum_{I,J}\cT_{IJ} \chi^I \bar\chi^J \ee
where $q=\exp(2\pi \i \tau)$ and the characters $\chi_I$ provide a
fully resolved unitary representation of the modular group. The
non-negative integers $\cT_{IJ}$ are tightly constrained by
modular invariance. Denoting by $I=0$ the character of the
identity representation of the RCFT, $\cT_{00}=1$ implies the
presence of only one graviton in the massless spectrum. Simple
solutions are: the charge conjugation modular invariant $\cT_{IJ}
= \cC_{IJ}$ (`Cardy'), and the `diagonal' modular invariant
$\cT_{IJ} = \delta_{IJ}$.

The massless spectrum is encoded in those combinations for which
$h_I = \bar{h}_J = 1/2$. Since $V_2$ already corresponds to $h_V =
1/2$, the only massless contribution of this kind comes from
$\cX^V_0$ which corresponds to the identity of the internal CFT.
Other massless bosons come from $O_2$ combined with $h_I^{int} =
\bar{h}_J^{int}= 1/2$. In Gepner models these are in one to one
correspondence with chiral (c) and anti-chiral (a) primary
operators with $q_I =\pm \bar{q}_J  = \pm 1$, respectively. In
type IIB, (c,c) states and their conjugate (a,a) states give rise
to $h_{2,1}$ $\cN=2$ vector multiplets, comprising two NS-NS
scalars and one R-R vector, while (c,a) states and their conjugate
(a,c) states give rise to $h_{1,1}+1$ $\cN=2$ hyper-multiplets,
comprising two NS-NS scalars and two R-R `axions' (dual to
two-forms). The special K\"ahler `geometry' of the vector
multiplets is tree level exact since corrections in $g_s = \langle
\phi \rangle$ are forbidden. Indeed the dilaton $\phi$ is part of
the universal hypermultiplet and as such it cannot have neutral
couplings to vector multiplets. The same argument applies to
worldsheet instanton corrections that depend on the sizes of the
holomorphic cycles governed also by scalars in hypermultiplets. On
the contrary, the dual quaternionic geometry of the
hypermultiplets can be corrected both perturbatively and
non-perturbatively.

The generalized $\Omega$-projection is encoded in the Klein bottle
amplitude
\be \cK = \sum_I K_I \chi^I (q\bar{q}) \ee
where $K_I = \cT_{II}$ ({\rm mod} 2) determines in particular
which massless fields are retained. Typically (but not
necessarily) both vector multiplets and hypers produce `chiral'
(or rather linear) multiplets. Yet if one splits $h_{2,1}$ into
$h^+_{2,1} + h^-_{2,1}$ where the apex indicates an extra possible
sign, constrained by the so-called crosscap constraint and
associated to some internal anticonformal involution, one can show
that the resulting unoriented spectrum contains $h^+_{2,1}$ chiral
multiplets and $h^-_{2,1}$ abelian vector multiplets, comprising
R-R vectors, in addition to $h^+_{1,1} + h^-_{1,1}$ chiral/linear
multiplets.

The open string partition function is given by
\be \cA = \sum_{I,a, b} A_{I a\bar{b}} n^a \bar{n}^b \chi^I  \ ,
\ee
where $n^a$ is the number of `generalized' D-branes of type $a$
and $A^I_{ab}$ are integer multiplicities constrained by the
quadratic equations
\be \sum_{b \bar b} A_{I a\bar{b}} \delta^{\bar b b} A_{J
b\bar{c}} = \sum_K N_{IJ}^{K} A_{K a\bar{c}} \ , \ee
where $N_{IJ}^{K}$ are the fusion rule coefficients, which can be
expressed in terms of the fully resolved $S_{IJ}$ via Verlinde
formula. Finally, the M\"obius strip $\Omega$-projection reads
\be \cM = \sum_{I,a,b} M_{I a} n^a \hat\chi^I  \ , \ee
where $M_{Ia} = A_{Iaa}$ (mod 2) and $\hat\chi^I$ denote a real
basis of characters introduced in \cite{Bianchi:1990yu}. We remind
that the arguments of the Annulus and M\"obius amplitudes are
different, namely $\tau_A = it/2$, $\tau_M = \tau_A + 1/2$. In
what follows (unless essential), we shall systematically omit the
$\tau$ dependence in the characters.

Worldsheet covariance conditions between the direct channel,
exposing the projection of the closed string spectrum ($\cK$)
or the open string spectrum ($\cA$ and $\cM$), and the transverse
channel (which is exposing the closed string exchange between
boundaries and crosscaps) puts tight constraints on the
coefficients $K_I$, $A_{I a\bar{b}}$ and $M_{I a}$.

For the case of the charge conjugation modular invariant $\cT_{IJ}
= \cC_{IJ}$ one has as many boundaries (\ie $n$'s)  as characters,
and one solution (known as Cardy's solution)
is given by
\be A_{IJK} = N_{IJK} \quad , \quad K_I = Y_{I00} \quad ,
\quad M_{IJ} = Y_{JI0} \, . \ee
Here $N_{IJK}$ are the fusion rule coefficients, while
$Y_{IJK}$ are (possibly negative) integers
given by
\be Y_{IJK} = \sum_L {S_{IL} P_{JL} P_{KL} \over S_{0L}} \, , \ee
where $P = T^{1/2} S T^2 S T^{1/2}$ is the M\"obius strip modular
matrix implementing the transformation $({\rm i} t +1)/ 2
\rightarrow ({\rm i} + t )/2 t$. The respective boundary and
crosscap reflection coefficients are
\be {\rm B}_I = {\sum_J S_{IJ} n^J \over \sqrt{S_{0I}}} \quad ,
\quad  \Gamma_I ={P_{0I} \over \sqrt{S_{0I}}}  \ . \ee

{}{}{}{}{}{}{}{}{}{}

\section{Tree level gauge couplings}

As suggested in \cite{Bianchi:2000de}, the `generalized'
Born-Infeld action for branes in (non)geometric backgrounds that
admit a (rational) CFT description can be extracted from
factorization of the one-loop annulus amplitude in the transverse
channel. This applies to Gepner models which are expected to
correspond to special (often non singular) points in the moduli
space of CY compactifications, where the K\"ahler and the complex
structure moduli take string scale VEV's \ie $R\approx \sqrt{\ap}$
and the supergravity approximation might be questionable. Yet, the
worldsheet string description is fully reliable in perturbation
theory. In fact non-perturbative effects in $R^2/\ap$ and even in
$1/g_s$ may be systematically incorporated.

\subsection{Tadpole cancellation and gauge couplings}

A consistent space time interpretation, requires the absence of
tadpoles for massless states, which schematically reads
\be {\rm B}_I + 2^{D/2} \Gamma_I = 0 \ , \qquad \forall I \ : \
h_I = 1/2 \quad . \ee
Although NS-NS tadpoles only signal an instability of the chosen
configuration, it has proved very hard to dispose of them by
vacuum redefinition \cite{FischSuss,DudNicoPradSag}. On the other hand, R-R
tadpoles are associated to anomalies \cite{Bianchi:2000de}.
In fact R-R tadpole conditions are more restrictive than simply
chiral anomaly cancellation that is associated to R-R tadpoles
in sectors with non-vanishing Witten index \cite{Bianchi:2000de}.
Actually some left-over anomalies involving $U(1)$ factors in the
Chan-Paton group can be disposed of by the combined effect of
axions, playing the role of St\"uckelberg fields, and generalized
Chern-Simons couplings \cite{Anastasopoulos:2006cz}. We will
henceforth assume that a solution to the R-R tadpole conditions
has been found, \ie a consistent choice of $n_a$ has been made.
Supersymmetry would then imply the absence of NS-NS
tadpoles~\footnote{As mentioned in the introduction, it is
sufficient that each pair of branes preserves some supersymmetry
in order for this to be true.}.

Tree level dependence of gauge couplings on massless closed string
moduli can be determined by considering a three-point amplitude on
the disk with one closed string insertion in the bulk and two
massless open string insertions (vector bosons) on the boundary.
The boundary is mapped to the brane $a$. The amplitude reads
\be \la cV^{(0)}_A(x_1) \int V^{(0)}_A(x_2) c\bar{c}
V^{(-1,-1)}_{ReZ}(z, \bar{z}) \ra \ee
for the CP even coupling and
\be \la cV^{(0)}_A(x_1) \int V^{(-1)}_A(x_2)
c\bar{c}V^{(-1/2,-1/2)}_{ImZ} (z, \bar{z}) \ra \ee
for the CP odd coupling. For the open string insertions one can
use the gauge boson vertex operators introduced previously. For
the closed string insertion one has to combine scalar vertex
operators for the Left and Right movers. Using $SL(2)$ invariance
one can put $z=\i$, $\bar{z}=-\i$ and $x_1=\infty$. Integration
over $x_2$ produces a constant and the overall factor is exactly
$B^I_a Tr_a(T_1 T_2)$ where the Chan Paton factor has replaced
$n_a$ that appears for an empty boundary. This measures
\be {\rm B}^I_a = \left. {\de f_a \over \de
Z^I}\right\vert_{Z_I=0} \ee
where we have assumed that the rational (Gepner) point corresponds
to $Z_I =0$ and 
\be f_a = {\i\vartheta_a\over 2\pi} + {4\pi \over g_a^2} \ ,\ee
is the gauge kinetic function for branes of type $a$.

One arrives at the above conclusion by `factorization' of
the one-loop non-planar amplitude in the transverse channel. If
$\chi_I$ is a massless character which starts with the complex scalar
field $Z_I$,
one can conclude that the tree-level gauge coupling is given by
\be
f_a (Z_I) = f_a (Z_I = 0) + {\rm B}^I_a Z_I \ ,
\ee
to lowest order in $Z_I$. In particular the dilaton dependence,
measuring the tension of the brane, is given by $n^a B^0_a Z_0$
where $Z_0=S$ to adhere to standard notation. In fact if $Z_I$
contains a pseudoscalar axion, shifting under some (gauged) PQ
symmetry, this is the full story, \ie $f$ is at most linear in
$Z_I$. This is always true in sectors with non-vanishing Witten
index. The dependence on $Z_I$ belonging to sectors with
vanishing Witten index can be more involved and they can appear
in the one-loop threshold corrections.
Moreover, multiplicities in sectors with ${\cN =1}$ susy \ie non
vanishing Witten index are excluded by our assumption that fixed point
ambiguities have been resolved. On the contrary, sectors with ${\cN =2}$
susy entail a twofold degeneracy at least. Scalars from sectors with
${\cN =4}$ susy can contribute to the tree level gauge
couplings but not to the one-loop thresholds. Anyway, it is remarkable
how a low-energy coupling can directly probe the structure of the
underlying RCFT coded in the $ {\rm B}^I_a$, that in turn depend
on the choice of $K_I$, $A_{Iab}$ and $M_{Ia}$ and the `resolved'
matrices $S_{IJ}$ and $P_{IJ}$. In particular the value of the
Weinberg angle at the string scale is related to the ratio of the
real parts of the gauge kinetic function for $SU(2)_W$ and the
properly normalized $U(1)_Y$,
\be \tan^2\vartheta_W = {g^2_Y \over g^2_W} = {\cR f_W \over \cR
f_Y} = {\cR {\rm B}^I_W Z_I \over \cR {\rm B}^J_Y Z_J} \ee
where as above $Z$'s runs over all closed string moduli fields
and, obviously, in order for the formula to be predictive at all,
one has to assume the closed string moduli have been stabilized by
some flux or non-perturbative effect.

{}{}{}{}{}{}{}{}{}{}

\section{One-loop thresholds corrections}

The purpose of this Section is to obtain explicit and (relatively)
simple formulae for the one-loop threshold corrections to the
gauge couplings. In four dimensions, gauge couplings run
logarithmically as a result of massless particles in the loops.
Massive states, such as generalized KK modes or genuine string
excitations, induce threshold corrections $\Delta_a$ in the form of
\be {1\over g_a^2(\mu)} = {1\over g_a^2(M)} + {b_a\over 8 \pi^2}
\log \left({\mu\over M}\right) + \Delta_a \ee
where $b_a$ is the coefficient of the one-loop $\beta$ function.
Threshold corrections signal the dependence on the light scalar fields
in the macroscopic theory of the mass scale $M$ at which the matching with the
microscopic theory is performed.

We will follow the strategy pioneered in \cite{Bachas:1992bh,
Bachas:1996zt} and successfully applied to type I orbifolds in
\cite{Antoniadis:1999ge,Bianchi:2000vb}, to generic  type I vacuum
configurations in \cite{Bianchi:2000vb} and to intersecting brane
models in \cite{Lust:2003ky}, based on the background field
method. We give only a very brief summary of the arguments. For
details see \eg \cite{Bianchi:2005sa}. The method consists in
applying a small abelian constant magnetic field in some spacetime
directions, computing the effect of such an integrable deformation
and then extracting the quadratic term in the one-loop effective
action.

Following \cite{Bianchi:2005sa}, we turn on an abelian magnetic
field in spacetime directions 2 and 3, leaving unmodified the
light cone directions 0 and 1,
\be F_{\m\n} = \delta_{[\m}^2 \delta_{\n]}^3 f \cH \ee
where $\cH$ is one of the generators  of the unbroken CP group.
Depending on the embedding of $\cH$ in the CP group one finds
different behaviors. To avoid complications we will focus only on
the case in which $\cH$ is a generator of a non-abelian and thus
non-anomalous factor labelled by $a$. Expanding the Annulus and
the M\"obius amplitudes ${\cal A}_a(f)$ and ${\cal M}_a(f)$ to
second order in $f$, one finds schematically for the one-loop gauge
threshold for the group to which belongs $\cH$
\cite{Bachas:1996zt,Antoniadis:1999ge}:
\be \Delta_a  =  \int \frac{dt}{t}({\cal A}_a''(0) + {\cal
M}_a''(0)) = \int \frac{dt}{4t} \cB_a(t) \, , \label{AMprime} \ee
where the prime denotes the derivative with respect to $f$. The
expression is IR divergent, signalling the running of the gauge
couplings, and needs regularization that, for non abelian gauge
groups simply amounts to replacing $\cB_a(t)$ with  $\cB_a(t) -
b_a$ with $b_a$ the on-loop $\beta$ function coefficient.

The presence of the magnetic field implies that the space-time
characters entering the Annulus and M\"obius amplitudes will have
a non-zero $z$ argument.
\be \chi_I(z,\tau) = V_2(z,\tau) \cX_{I}^{V} (0,\tau) +
O_2(z,\tau) \cX_{I}^{O} (0,\tau) - S_2(z,\tau) \cX_{I}^{S}
(0,\tau) - C_2(z,\tau) \cX_{I}^{C} (0,\tau) \, , \label{ychar1}
\ee
where $I$ labels the different orbits / sectors in the theory. Let
us stress that $z \neq 0 $ ($f \neq 0$) in eq.~(\ref{ychar1})
breaks supersymmetry, so the characters $\chi_I(z,\tau)$ are not
identically zero anymore. The second derivative with respect to
$f$ in eq. (\ref{AMprime}) translates into a second derivative
with respect to $z$ of the characters $\chi_I(z,\tau)$. Since only
the space time is $z$ dependent, one finds
\be \cB_I(0) = V_2''(0) \cX_I^V (0) + O_2''(0) \cX_I^O (0) -
S_2''(0) \cX_I^S (0) - C_2''(0) \cX_I^C (0) \, , \ee
where here, and in the rest of this Section, prime denotes a
derivative in $z$. Putting all pieces together
\be \cB_a(t) = \sum_{I,b}A^I_{ab} n^b \cB_I(t) + \sum_{I} M^I_{a}
\hat\cB_I(\hat t) \, ,  \ee
where $A^I_{ab}$,
$M^I_{a}$ are integer multiplicities and $ n^b$ are the number of
branes in each stack.

{}{}{}{}{}{}{}{}{}{}

\subsection{Thresholds from $\cN=2$ SCFT}

Due to  the very complicated form of the internal characters
$\cX^\lambda_I$ in Gepner models, the above expression for
$\cB_I(0)$ is not very useful. In order to rewrite it in a more
tractable form, let us introduce the supersymmetric $SO(2)\times
U(1)_R$ spacetime characters $v,\phi, \phi^\dagger$ defined by
\cite{Lerche, Angelantonj:1996uy}
\bea v(z,y) &=& V_2(z) \xi_0(y) + O_2(z) \xi_3(y) - S_2(z)
\xi_{+3/2}(y)- C_2(z) \xi_{-3/2}(y) \, ,  \nonumber \\
\phi(z,y) &=& V_2(z) \xi_{-2}(y) + O_2(z) \xi_{+1}(y) - S_2(z)
\xi_{-1/2}(y) - C_2(z) \xi_{+5/2}(y) \, , \label{vphiphi} \\
\phi^c(z,y) &=& V_2(z) \xi_{+2}(y) + O_2(z) \xi_{-1}(y) - S_2(z)
\xi_{-5/2}(y) - C_2(z) \xi_{+1/2}(y) \, . \nonumber \eea
Here $\xi_{p}(y)$ which encode the coupling to the total
R-symmetry charge $J_R = \sum_i J_R^{(i)}$, are given by
\be \xi_{p}(y) = {1\over \eta} \sum_n q^{{1\over 6} (p + 6n)^2}
e^{2\pi \i y (p + 6n)}  \ee
and satisfy
\be 12 \pi {\rm i } \,  \de_\tau \left( \xi_{p}(y) \eta \right)  =
\eta \, \de^2_y \xi_{p}(y)  \, . \label{d2xi} \ee

In any SUSY compactification to $D=4$, the characters can be
decomposed according to
\be \chi_I (z,y,\tau) = v(z,y,\tau) \hat\chi^v_I(\tau) +
\phi(z,y,\tau)  \hat\chi^\phi_I(\tau) + \phi^c(z,y,\tau)
\hat\chi^{\phi_c}_I(\tau) \, , \label{chizy} \ee
where $\hat\chi^\Lambda$ are characters of $(\cN = 2)/U(1)_R$. It
is quite remarkable and crucial for our subsequent analysis that
\be v(z,y=z/3) =0 \, , \qquad \phi(z,y=z/3) =0 \,  , \qquad
\phi^c(z,y=z/3) =0 \, , \ee
for any  $z$ thanks to theta functions identities ({\it cf. e.g.} \cite{Sarkissian:2006fn}). Then it follows immediately that also
\be \chi_I (z,z/3,\tau) = 0 \, , \label{chizz/3} \ee
for all values of $I$, $z$ and $\tau$. This tantalizingly suggest
the possibility of building more general supersymmetric
`magnetized' aka `coisotropic' branes. We will come back to this
issue in a later section.

Taking the first derivative with respect to $\tau$ and the second
derivative with respect to $z$ of eqs.~(\ref{vphiphi}) for $y=z/3$
and using that $4 \pi {\rm i } \de_\tau \chi^{SO(2)}_\lambda(z) = \de_z^2
\chi^{SO(2)}_\lambda(z)$ (up to an irrelevant $\eta$) as well as
eq.~(\ref{d2xi}) one finds
\be \cB + {1 \over 3} \cA  = 0 \ , \qquad \cB +  {1 \over 9} \cA =
- {2 \over 3} \cC \ . \ee
Here $\cB$ collectively denotes terms with second derivative of
$\chi^{SO(2)}_\lambda(z)$ (\ie which contribute to the
thresholds), $\cA$ denotes terms with second derivative of
$\xi_p(z/3)$ and $\cC$ terms with two first derivatives.
Eliminating $\cA$, one finds $ \cB = - \cC $ and after
substituting in eqs.~(\ref{chizy},\ref{vphiphi}) one then gets
\bea \cB_I(z, z/3) = &-& [ V_2'(z) (\cX_I^{V})' (z/3)
+ O_2'(z) (\cX_I^{O})' (z/3) \nonumber \\
&& - S_2'(z) (\cX_I^{S})' (z/3) - C_2'(z) (\cX_I^{C})' (z/3)] \, ,
\eea
where $\cX_I^\lambda(z/3)$ denotes the character valued internal
partition function in the relevant sector of the orbit $I$.

For $z=0$, $V_2'(0) = O_2'(0) = 0 $, while $S_2'(0) = - C_2'(0) =
\i {\theta_1}'/2  =  \i \pi \eta^3$ that cancels a similar factor
in the denominator. So finally we obtain

\be\label{cs} \cB_I =  \left. {d \over dy}(\cX_I^{S} (y) - \cX_I^{C}
(y)) \right|_{y=0} \, . \ee
We stress that this general formula is  valid for any  susy
compactification to $D=4$.
In the particular case of Gepner models,
it can be additionally simplified. Indeed, using eqs.~(\ref{ChiS})
and (\ref{ChiC}),
we can write
\be \cB_I = \left. {d \over dy} \cW_I(y) \right|_{y=0} \, ,
\label{treshD} \ee
where $\cW_I(y)$  in the sector $I$ is given by
\be \cW_I(y) =  (-1)^{r + \sum_i {m_i \over k_i+2}} \
 \sum_{n=0}^{K/2-1} (-1)^{n(r-1)} \prod_{i=1}^r
\cW^{\ell_i}_{m_i- 2n-1}(y) \,  , \ee
where
\be \cW^{\ell_i}_{m_i-2n-1}(y) = \chi^{\ell_i}_{m_i- 2n-1, 1}(y) -
\chi^{\ell_i}_{m_i- 2n-1, -1}(y) = Tr_{{\cal H}_{i,n}}[ (-)^F
e^{2\pi \i y J_o} q^{L_o - c_i/24} ] \ee
is called elliptic index. For $y=0$ it is a constant, and since
only the Ramond groundstates contribute
\be \cW^{\ell_i}_{m_i- 2n-1}(y=0) = \cI^{\ell_i}_{m_i-2n-1} =
\delta_{m_i - 2n-1, \ell_i + 1} - \delta_{m_i-2n-1, -\ell_i - 1}
\, , \label{cindex} \ee
where both deltas are computed mod $2(k_i+2)$. Thus the derivative
in eq.~(\ref{treshD}) reduces to
\be {\cW_I}'(y=0) =   (-1)^{r + \sum_i {m_i \over k_i+2}} \
 \sum_{n=0}^{K/2-1} (-1)^{n(r-1)} \sum_{j=1}^r
(\cW^{\ell_j}_{m_j- 2n-1})'(y=0) \prod_{i=1 \atop i \neq j }^r
\cI^{\ell_i}_{m_i- 2n-1} \,  . \ee
This expression can be further simplified with the help of eq.~(\ref{cindex}).

Starting from the expression for $\chi_{l,m}^{NS^+}$,
given in  \cite{Aldazabal:2003ub}, one can derive the expression
for $ \cW^\ell_m = \chi_{l,m}^{R^-}$ by a shift of the argument
$z \rightarrow z+(\tau+1)/2$
\bea \cW^\ell_m(z)&=&\chi^\ell_{m,1}(z)-\chi^\ell_{m,-1}(z)\\
\nonumber && \nn \cr
&=&\frac{e^{\i\pi\left(\frac{\ell+m+1}{k+2}-1\right)} \
 \theta_1(z,\tau) \ \theta\left[{-\frac{\ell+1}{k+2}
+\frac{1}{2} \atop \frac{1}{2} }\right](0,(k+2)\tau) \
\eta^3((k+2)\tau)} { \eta^3(\tau) \
\theta\left[{\frac{\ell+m+1-(k+2)}{2(k+2)}\atop
\frac{1}{2}}\right](z,(k+2)\tau) \ \theta \left[
{\frac{-\ell+m-1+(k+2)}{2(k+2)}\atop
\frac{1}{2} }\right](z,(k+2)\tau)} \nn\\
&& \nn \cr && \nn \cr
&=&\frac{e^{-\i\pi z\left(\frac{m}{k+2}\right)}~
 q^{\frac{(\ell+1)^2-m^2}{4(k+2)}}\eta^3\Big((k+2)\tau\bigg) ~\theta_1(z,\tau)~
\theta_1\bigg((\ell+1)\tau,(k+2)\tau\bigg)}
{
\eta^3(\tau)~\theta_1\bigg(z-\frac{\ell-m+1}{2}\tau,(k+2)\tau\bigg)
~ \theta_1\bigg(z+\frac{\ell+m+1}{2}\tau,(k+2)\tau\bigg)} \, .
\label{Wconj}
 \eea
It is immediate that $\cW^\ell_m(0)=0$ unless $m=\ell+1$ or
$m=-(\ell+1)$, $\cW^\ell_{\ell+1}(0)=1$ and
$\cW^\ell_{-\ell-1}(0)=-1$. Moreover one can show that
\be \sum_{\ell=0}^k\cW^\ell_{\ell+1}(z)=\frac{
\theta_1(\frac{k+1}{k+2} z,\tau)} {\theta_1(\frac{1}{k+2} z,\tau)}
\ . \ee Let us denote \be
 a={\ell+1 \over (k+2)} \, , \quad
 b={\ell+1-m \over 2(k+2)} \, ,  \quad
 c={\ell+1+m \over 2(k+2)} \, .
\label{abc} \ee Then for the derivatives $(\cW^\ell_m)'(0)$ one
finds \be (\cW^\ell_{\ell+1})'(0) = (\cW^\ell_{-\ell-1})'(0) = {d
\over dy} \left. \ln \left( \theta\left[{{1 \over 2} - a \atop
\frac{1}{2}}\right](y,(k+2)\tau) \right) \right|_{y=0} \, ,
\label{Wmassless} \ee while if $m \neq \ell+1, -\ell-1$ \be
(\cW^\ell_{m})'(0) = 2 \pi \i \, q^{(\ell+1)^2-m^2 \over 4(k+2)}
{\cP_k(0)^2 \ \cP_k(a)  \  \cP_k(1-a) \over \cP_k(b) \ \cP_k(1-b)
\ \cP_k(c) \ \cP_k(1-c)} \, , \label{Wmassive1} \ee where \be
\cP_k(\alpha) = \prod_{n=1}^{\infty} (1-q^{(k+2)(n-\alpha)}) \, .
\label{Wmassive2} \ee

{}{}{}{}{}{}{}{}{}{}

\subsection{Thresholds in toroidal orbifolds}

For completeness and for comparison, let us summarize here known
formulae for the thresholds corrections to gauge couplings in Type
I (magnetized) toroidal orbifolds. It is known that some Gepner models, \eg
$(k=1)^9$ or $(k=2)^6$ models in $D=4$,
correspond to toroidal orbifolds at special points in their moduli spaces.
Formulae in this section would then apply to these cases.
For brevity we only discuss the
contribution of ${\cal N}=1$ supersymmetric sectors.  Expanding
the annulus and M\"obius strip amplitudes to quadratic order in
the background field $f$ and summing over spin structures by means
of
\be \sum_{\a\b} c_{\a\b}\frac{\theta''[^\a_\b](0)}{\eta^3} \prod_I
\frac{\theta[^\a_\b](u^I)} {\theta_1(u^I)} = 2\pi \sum_I
\frac{\theta_1'(u^I)} {\theta_1(u^I)}  \ , \ee
give
\bea\label{Ba} \cB_a^{{\cal N}=1}(t)  & = & \frac{i}{\pi} \sum_{b}
\cI_{ab} N_b \sum_I \frac{\theta_1'(u^I_{ab}|\tau_A)}
{\theta_1(u^I_{ab}|\tau_A)} \nonumber\\ \hat{\cB}_{a}^{{\cal
N}=1}(t)  & = & -\frac{i}{\pi} \sum_{a} \cI_{a\tilde a} N_a\sum_I
\frac{\theta_1'(u^I_{a\tilde a}|\tau_M)} {\theta_1(u^I_{a\tilde
a}|\tau_M)} \ , \eea
where
\be u^I_{ab} = \kappa v^I_{ab} + \epsilon^I_{ab}\tau  \ ,\ee
satisfy $\sum_I u^I_{ab} = 0$ and take into account both the
orbifold projection $\kappa v^I_{ab}$ (\eg $\kappa = 1,..., n$ for
$\Gamma = {\bf Z}_n$) and the mass shift
$\epsilon^I_{ab}$ due to magnetic flux or intersections at angle.
The one-loop $\b$-function coefficients can be extracted from the
IR limit of (\ref{Ba}).

In order to perform the integral and compute $\Delta_a$ in
magnetized tori ($v^I_{ab} =0$), it is convenient to switch to the
transverse channel, where one finds
\bea\label{thr} \Delta_{a}^{{\cal N}=1} & = &
\frac{1}{2 \pi}\sum_{a,b} \cI_{ab} N_b
\sum_I\int_{0}^{\infty}\frac{\theta_1'(\epsilon^I_{ab}|i\ell)}
{\theta_1(\epsilon^I_{ab}|i\ell)}\;d\ell  \ ,\nonumber\\
\hat\Delta_{a}^{{\cal N}=1} & = & -\frac{1}{2\pi} 2 \cI_{a\tilde
a} \sum_I\int_{0}^{\infty}\frac{\theta_1'(\epsilon^I_{0a}|i\ell +
1/2)} {\theta_1(\epsilon^I_{0a}|i\ell + 1/2)}\;d\ell \ .\eea
Series expansion
\be \frac{{\theta_1'(\epsilon|\tau)}}{\theta_1(\epsilon|\tau)} \pi
\cot(\pi\epsilon) + 2 \sum_{k=1}^\infty \zeta(2k) \epsilon^k
(E_{2k}(\tau) - 1) \quad ,\ee
where $ \zeta(2k) = (2\pi)^{2k}|B_{2k}| / (2k)!$ and
$E_{2k}(\tau)$ is an Eisenstein series with modular weight $2k$,
expose potentially divergent terms that eventually cancel thanks
to (NS-NS) tadpole cancellation, for the non-anomalous $\cH$, with
$Tr(\cH)=0$. The finite terms boil down to integrals of the form
\bea &&\int_0^\infty d\ell \sum_{k=1}^\infty 2\zeta(2k) \epsilon^k
(E_{2k}(i\ell) - 1) = - \pi\log\left[\Gamma(1 - \epsilon) \over
\Gamma(1 + \epsilon)\right] + 2\pi\epsilon \gamma_E\ ,\\
\label{threshintegral}
&&\int_0^\infty d\ell  \sum_k 2\zeta(2k) \epsilon^k (E_{2k}(i\ell
+ 1/2) - 1) = - \pi\log\left[\Gamma(1 - 2\epsilon) \over \Gamma(1
+ 2\epsilon)\right] + 2\pi\epsilon \gamma_E \ .\ \ \ \ \eea
Actually the last contributions, linear in $\epsilon$, drop after
summing over the three internal directions in supersymmetric
cases.

Summing the various contributions one finally gets
\bea\Delta^{{\cal N}=1}_{a} & = & - \sum_{b} \cI_{ab}N_b \sum_I
\log\left[\Gamma(1 - \epsilon^I_{ab}) \over \Gamma(1 +
\epsilon^I_{ab})\right]  \quad  , \nonumber\\
\hat\Delta^{{\cal N}=1}_{a} & = & \sum_{a} 2 \cI_{a \tilde a}
\sum_I \log\left[\Gamma(1 - \epsilon^I_{aa}) \over \Gamma(1 +
\epsilon^I_{aa})\right]  \quad ,\eea
where $\epsilon^I_{aa} = 2 \epsilon^I_{ao}$.

Field dependent thresholds corrections from ${\cal N}=2$ sectors  with
vanishing Witten index ($u^I_{ab}=0$ for
some $I=\parallel$, so that $u^{\perp, 1}_{ab}=-u^{\perp,
2}_{ab}$)  are much easier to compute since they correspond to BPS
saturated couplings. We refrain from doing so explicitly here. ${\cal N}=4$
sectors ($u^I_{ab}=0$ for all $I$) do not contribute threshold
corrections to the gauge couplings.

{}{}{}{}{}{}{}{}{}{}

\section{Examples}

Once the general formula has been derived, in order to  compute
explicit thresholds one has to put together various bits and
pieces.

First one has to fix the integer multiplicities in the annulus and
Moebius amplitudes compatibly with tadpole cancellation.

Second one has to choose a non-abelian group and identify  the
sectors of the open string spectrum which are charged. We neglect
possibly anomalous $U(1)$'s since the above formulae do not
immediately apply. In fact they rather compute the masses of the
gauge bosons via their mixings with R-R axions.

Third one has to perform the integral over $t$. This was  done
above for magnetized tori and it is possible for some contributions
(from fully massless sectors) in type I Gepner models as well.

Let us discuss what happens in various dimensions.

\subsection{Models in $D=8$}

In $D=8$ supersymmetric models correspond to compactifications on
2-tori. The two derivative effective action is tree level exact
because of susy. Some four derivative terms such as $F^4$ are 1/2
BPS  saturated. Starting from the seminal paper by Bachas and
Fabre \cite{Bachas:1996zt}, one-loop threshold corrections to
these and other BPS saturated have been used as tests of various
string dualities. For a comprehensive review see
\cite{Eliasreview}.

\subsection{Models in $D=6$}

Threshold corrections in $D=6$ are topological in the sense  that
only massless states can contribute. Indeed in theories with $\cN
= (1,0)$ susy the gauge couplings can only depend on the VEV's of
scalar that belong to tensor multiplets and not to hypermultiplets
because of susy. In perturbative heterotic models the only tensor
multiplet contains the dilaton and this produces the standard
dependence of the gauge coupling from the string coupling. All the
remaining moduli,  either charged or neutral, belong to
hypemultiplets. In type I constructions \cite{Bianchi:1990yu}
various neutral tensor multiplets are present whose scalar
components belong to the NS-NS sector. In principle gauge
couplings may depend on them. There is a tight connection with anomaly
related couplings as required by the generalized mechanism of
anomaly cancellation.

After compactification to $D=4$ on a 2-torus one gets $\cN = 2$
theories whose gauge kinetic function is 1/2 BPS saturated. Only
generalized KK modes contribute to the threshold. Generalized
compactifications \'a la Scherk-Schwarz with freely acting
orbifolds preserving $\cN = 1$ may lead to interesting
applications of our analysis in connection with large extra
dimensions.

\subsection{Models in $D=4$. The Quintic : $(k=3)^5$ model}

The simplest non trivial case is a Type I model on the  Quintic
\cite{Blumenhagen:1998tj,Blumenhagen:2003su,Aldazabal:2003ub}. It is
based on the diagonal modular invariant that puts fewer tadpole
constraints than the charge conjugation modular invariant. Indeed,
in the transverse channel only two massless sectors can propagate.
The identity and the sector containing the unique (c,a) massless
state (unique deformation of the K\"ahler structure). To cancel
tadpoles one can introduce so-called B-type branes and in
particular one can build a model with $SO(12)\times SO(20)$
Chan-Paton group. Though non chiral, the model serves as a non
trivial illustration of our procedure.

The annulus partition function is given by
\be {\cal A} = {1\over 2}
(n_0^2 + n_1^2) \chi_A + ({1\over 2} n_1^2 + n_0 n_1) {\chi_B} \ ,
\ee
where $n_0 =12$ and $n_1 =20$.
The M\"obius strip projection reads
\be {\cal M} = - {1\over 2} (n_0 +
n_1) \hat\chi_A + {1\over 2} n_1 \hat{\chi}_B \ .
\ee
Here ${\chi_A}$ and ${\chi_{B}}$ are given by
\bea
\chi_A &=& {1 \over 5}  \left[ (\chi_{I})^{5} \right]^{\rm susy} \ , \nonumber \\
\chi_B &=& {1 \over 5} \left[ (\chi_{I})^{4} \, \chi_{II} \right]^{\rm susy} \ ,
\eea
where $\chi_I$ and $\chi_{II}$ are defined as
\bea
\chi_{I} &=& {1 \over 2}
\left(
\chi^{0}_{0,0}+\chi^{0}_{0,2}+\chi^{0}_{2,0}+\chi^{0}_{2,2}+\chi^{0}_{4,0}+\chi^{0}_{4,2}
+\chi^{0}_{6,0}+\chi^{0}_{6,2}+\chi^{0}_{8,0}+\chi^{0}_{8,2} \right) \, , \\
\chi_{II} &=& {1 \over 2} \left(
\chi^{1}_{1,0}+\chi^{1}_{1,2}+\chi^{1}_{3,0}+\chi^{1}_{3,2}+\chi^{1}_{5,0}+\chi^{1}_{5,2}
+\chi^{1}_{7,0}+\chi^{1}_{7,2}+\chi^{1}_{9,0}+\chi^{1}_{9,2}
\right) \, ,
\eea
in terms of the $\cN=2$, $k=3$ characters $\chi^{\ell}_{m,s}$.

The massless spectrum is given by $\cN =1$ vector multiplets in
$Adj[SO(20)\times SO(12)]= ({\bf
190} + {\bf 66})$ plus four chiral multiplets in the
$({\bf 20},{\bf 12})$ and as many in the $({\bf
210}, {\bf 1})$.
One can thus easily compute the $\beta$ functions for $SO(20)$ and
$SO(12)$ and get  \bea &&\beta_{SO(20)} = 3 (20-2) - 4 (12 +
(20+2)) = - 82
\nn \\
&&\beta_{SO(12)} = 3 (12-2) - 4 (20) = - 50 \eea both gauge
couplings are IR free.

From eqs.~(\ref{treshD}-\ref{Wconj}) we find: \be {\cB_A} = - 5
(\cW^0_{1})' - 5 (\cW^0_{9})' - 30 (\cW^0_{5})' + 20 (\cW^0_{3})'
+ 20 (\cW^0_{7})' \ , \ee and
\bea
{\cB_B}  =
&& 8 (\cW^0_{1})' +8 (\cW^0_{9})'  - 4 (\cW^1_{2})' -4 (\cW^1_{8})'  + \nn \\
&& 8 (\cW^0_{5})'  -12  (\cW^0_{3})'  - 12  (\cW^0_{7})'
+ 6  (\cW^1_{0})'   + (\cW^1_{4})' + (\cW^1_{6})' \ .
\eea
These derivatives can be computed with the help of
eqs.~(\ref{abc}-\ref{Wmassive2}). In particular for the
contributions relevant for the $\beta$ functions are
\bea
(\cW^0_{1})'= (\cW^0_{9})' &=& {3 \over 5} \i \pi + \dots \\
(\cW^1_{2})'= (\cW^1_{8})' &=& {1 \over 5} \i \pi + \dots \eea
Contributions of fully massless sectors can be computed by means of
(\ref{Wmassless})
and integrated by means of (\ref{threshintegral}). The contributions to the
thresholds that
involve one massive subsector can be computed by means of (\ref{Wmassive1}).
We have not yet been able to find a simple way to integrate the result as for
the fully massless sectors.

{}{}{}{}{}{}{}{}{}{}

\section{Magnetized aka coisotropic D-branes}
\label{magncoiso}

In toroidal or orbifold compactifications  one can easily impose
`generalized' boundary conditions that correspond to turning on a
constant magnetic field on the worldvolume of the D-brane
\be [\de X^i - R_a^{i}{}_{j} \bar\de X^j]|a\ra_F = 0 \quad , \quad
[\psi^i -\i \eta R_a^{i}{}_{j} \bar\psi^j]|a\ra_F = 0 \ee
where $\eta=\pm 1$, depending on the sector, and the orthogonal matrix (in the
frame basis) reads
\be R_a^{i}{}_{j} = [\delta_a^{i}{}_{k} - F_a^{i}{}_{k}]
[\delta_a^{k}{}_{j} + F_a^{k}{}_{j}]^{-1} \ee $R_a^{i}{}_{j}$
can be diagonalized in a complex $a$-dependent basis $Z^I, Z^*_I$,
so that
\be \de Z^I = e^{2\pi \i \nu_a^I} \bar\de Z^I \ee
as a result the modes of $Z^I$ are shifted according to $ n^I
\rightarrow n^I + \nu_a^I$. A similar analysis applies to the
complex fermions $\Psi^I, \Psi^*_I$ such that $G = \de Z^*_I\Psi^I
+ \de Z^I\Psi^*_I$ ($a$-independent!!). When several stacks of
magnetized branes are present, the rotation matrices $R_a$ and
$R_b$ for different stacks would not commute in general. When
$[R_a,R_b] = 0$ for all $a$ and $b$, all the magnetic fields are
{\it parallel}, otherwise $[R_a,R_b]\neq 0$ and the magnetic
fields are {\it oblique}. Performing appropriate T-dualities on
magnetized D9-branes one ends up with intersecting magnetized
D-branes aka coisotropic D-branes. For parallel fields appropriate
T-dualities lead to intersecting D-branes with no magnetization
aka isotropic branes.

We would like to extend this analysis to compactifications  based
on genuinely interacting $\cN = 2$ SCFT.

For simplicity one can consider Gepner models first. In this case
the worldsheet supercurrent is given by
\be G = \sum_i [\psi^{PF}_i e^{\i\sqrt{3\over c_i} H_i} +
\psi^{PF,\dagger}_i e^{-\i\sqrt{3\over c_i} H_i}] \quad . \ee
There are two classes of boundary conditions preserving the
diagonal $\cN = 2$ SCA commonly called of A and B type. A-type
boundary conditions imply
\be [\psi^{PF}_i - \i \eta \bar\psi^{PF,\dagger}_i]|b\ra_A = 0 \ ,
\quad [e^{\i\sqrt{3\over c_i} H_i} - e^{- \i\sqrt{3\over c_i}
\bar{H}_i}]|b\ra_A = 0 \ee
and correspond to D-branes wrapping middle homology cycles (\ie
Special Lagrangian submanifolds) or generalized bound-states
thereof.

B-type boundary conditions imply
\be [\psi^{PF}_i - \i \eta \bar\psi^{PF}_i]|b\ra_B = 0 \ , \quad
[e^{\i\sqrt{3\over c_i} H_i} - e^{ \i\sqrt{3\over c_i}
\bar{H}_i}]|b\ra_B = 0 \ee
and correspond to D-branes wrapping even-dimensional  homology
cycles (\ie complex submanifolds) or generalizations thereof.

One can envisage the possibility of imposing symmetry breaking
boundary conditions such as
\be [\psi^{PF}_i - \i \eta e^{2\pi \i \nu_b^i}\bar\psi^{PF,
\dagger}_i]|b\ra_{\tilde A} = 0 \ , \quad [e^{\i\sqrt{3\over c_i}
H_i} - e^{-2\pi\i \nu_b^i} e^{- \i\sqrt{3\over c_i}
\bar{H}_i}]|b\ra_{\tilde A} = 0 \ee or  \be [\psi^{PF}_i - \i \eta
e^{2\pi \i \nu_b^i}\bar\psi^{PF}_i]|b\ra_{\tilde B} = 0 \ , \quad
[e^{\i\sqrt{3\over c_i} H_i} - e^{-2\pi \i \nu_b^i} e^{
\i\sqrt{3\over c_i} \bar{H}_i}]|b\ra_{\tilde B} = 0 \ee
that should naturally correspond to D-branes wrapping submanifolds
with non trivial magnetic fluxes and thus would deserve the name
of `coisotropic' D-branes in this context. More pragmatically the
boundary conditions combine a shift in the $U(1)$ charge lattice
with a compensating `rotation' of the complex parafermions so as
to preserve the diagonal $\cN = 2$ SCA. In cases where several
factors are isomorphic (\ie have the same $k$) additional
`permutations' are possible in the boundary conditions leading to
what have been called `permutation' branes. The open string
excitations of this more general class of D-branes belong to
twisted representations of $\cN = 2$ SCA that are known to exist
for any real values of $\nu_b^i$. Spacetime supersymmetry imposes
further constraints \cite{Mizoguchi:2001xi,Sarkissian:2006fn}. A detailed study of this class of branes is
deferred to future work. Suffice it to say that including this new
class of branes enormously widens the possibilities of accomodating
interesting chiral models in Type I Gepner models.

For the time being let us check the validity of the above
interpretation for the phenomenologically uninteresting case of
$D=8$, \ie to ${\bf T}^2$ compactifications \cite{Giveon:1990ay,Gutperle:1998hb}, where a precise
dictionary exist between the standard bosonic and fermionic
coordinates $X, \psi$ and parafermions $\psi^{PF}$ and free boson
$H$. Indeed for the $(1,1,1)$ model $c=1+1+1=3$ and $H = \sum_i
H_i/\sqrt{3}$ and
\be \Psi = e^{\i H}~, \quad \de Z = {1\over \sqrt{3}} \sum_i
e^{\i(H-\sqrt{3}H_i)} \ee
while for the $(2,2,0)$ model $H = \sum_i H_i/\sqrt{2}$
\be \Psi = e^{\i H}~, \quad \de Z = {1\over \sqrt{2}} \sum_i \psi_i
e^{\i(H-2H_i)}~. \ee
Finally, for the $(4,1,0)$ model  $H = (\sqrt{2}H_1 +
H_2)/\sqrt{3}$
\be \Psi = e^{\i H}~, \quad \de Z = {1\over \sqrt{2}}
\left[\hat{\Psi}_{3/4} e^{\i(H- \sqrt{2\over 3}H_1)} + e^{\i(H-
\sqrt{3}H_2)}\right]~. \ee
Switching on  a non vanishing
$\nu_b^i \neq 0$ is tantamount to turning on a magnetic field or,
equivalently after T-duality, rotating the brane wrt the
fundamental cell of the ${\bf T}^2$.

{}{}{}{}{}{}{}{}{}{}

\section{Concluding remarks}

We have derived very compact and elegant formulae that allow  one
to determine the tree level gauge couplings and the one-loop
thresholds in Type I or similar compactifications based on
genuinely interacting $\cN =2$ SCA, such as Gepner models but not
only. We have then given some explicit example for the non-abelian
factors in the Chan-Paton gauge group. In view of
\cite{Anastasopoulos:2006cz,Anastasopoulos:2007qm} the analysis of anomalous $U(1)$
factors may reserve for us new interesting possibilities. Moreover
the computation of four vector boson scattering amplitudes at
one-loop seems at reach, since the threshold encode the structure
called $\cE$. The other irreducible
structure $\cF$ require some more work.The analysis
might be significantly simplified resorting to the hybrid formalism proposed
by Berkovits\footnote{M.B. would like to thank Nathan Berkovits for clarifying
discussions on this issue.}

We have then briefly discussed how to generalize the standard
boundary conditions so as to describe magnetized aka coisotropic
D-branes. This new class of branes may open new paths not only to
the construction of viable Type I models but also to the
generation of non-perturbative effects, \ie D-brane instantons,
mediated by magnetized or coisotropic ED-branes. It is in fact
more than natural to expect that ED-branes wrapping the same cycle
as a given stack of branes, including magnetization, are equivalent to standard gauge instantons
for the resulting effective theory, while all
other ED-branes generate stringy non perturbative phenomena.

Clearly before even contemplating stringy instanton effects  in
these backgrounds one should reliably compute tree level Yukawas
and K\"ahler potential for the open string excitations. We hope to
report on these issues soon although the perspectives of making
reasonable predictions for the Cabibbo angle in this context are much
weaker than for the Weinberg angle.
It would also be interesting to study models with large extra
dimensions \'a la Aldazabal et al \cite{Aldazabal:2004by,Aldazabal:2006nz}or even non-susy models with
supersymmetric partition functions. As mentioned in the
introduction the final goal would be to stabilize all moduli and
break susy in a controllable way. This may not forgo understanding
better, from a worldsheet vantage point the effects of fluxes and
gaugings.

\section*{Acknowledgements}

We would like to thank Bert Schellekens for e-mail correspondence.
Some of the basic ideas in this paper were presented by M.B. at the Indian
Strings Meeting 2006, Puri, India. M.B. would like to thank the organizers,
especially A. Kumar and
S. Mukherji, for creating a stimulating atmosphere and the participants,
in particular N. Berkovits, A. Dabholkar, S. Govindarajan, C. Hull, A. Kumar,
S. Mukhopadhyay and K. Ray, for interesting discussions.
During the completion of this work M.B. was visiting IOP in Bhubaneswar and
IACS in Kolkata, while P.A. was visiting  Ecole Polytechnique
in France,  that are acknowledged for their kind hospitality.
This work was supported in part by INFN, by the MIUR-COFIN
contract 2003-023852, by the EU contracts MRTN-CT-2004-503369 and
MRTN-CT-2004-512194, by the INTAS contract 03-516346 and by the
NATO grant PST.CLG.978785.

\appendix

\renewcommand{\theequation}{\thesection.\arabic{equation}}
\addcontentsline{toc}{section}{Appendix}
\section*{Appendix}

\section{The $k=3$,  $\cN =2$ minimal model}

In order to work out the thresholds for the example of the quintic
described by the $(k=3)^5$ Gepner models it is helpful to
decompose the primaries of the $k=3$ minimal model ($c=9/5$) into a
$U(1)$ model ($c=1$) combined with the 3 state Potts model, ($c=4/5$).
The primaries of the $U(1)$ model  are $V_q = \exp(\i q \sqrt{5/3}H)$
with $h= 5q^2/6$, where   $q$ is the charge. In the NS sector $q = n/5 = 2n/10$,
while in the R sector $q = (2n+1)/10$.
The primaries of the 3 state Potts model (which is actually a
quotient of the $c=4/5$ minimal model wrt a spin 3 W symmetry) are
six: $I$  identity  with $h=0$, $\epsilon$ energy with $h=2/5$ (real),
$\sigma$ and $\sigma^*$ spins with $h=1/15$, $\rho$ and $\rho^*$
parafermions with $h=2/3 =(k-1)/k$. Indeed the S-modular transformation
reflects the $Z_3$ symmetry $S = S_3 \otimes S_2$
where $S_3$ is the $S$-matrix of $SU(3)$ at level 1 and
\bea
S_2 = \frac{2}{\sqrt{5}} \left(\begin{array}{cc} s_1& s_2 \\
s_2& -s_1\end{array} \right)\eea
where $ s_n = \sin(n\pi/5)$.
The resulting fusion rules also reflect this symmetry.
In particular, $I$, $\rho$ and $\rho^*$ are simple currents. The only non obvious ones are
\be
\rho \times \sigma = \sigma^* \quad, \quad \rho \times \epsilon = \sigma \quad, \quad \rho \times \sigma^* = \epsilon
\ee
and their conjugate,
while $\epsilon$,
$\sigma$ and $\sigma^*$ have non abelian (`minimal' in a sense) fusion rules
\be
\epsilon \times \epsilon = \sigma\times \sigma^* = I + \epsilon
\ee
as well as
\be
\epsilon \times \sigma = \sigma^* \times \sigma^* = \rho + \sigma
\ee
and its conjugate.

In the Table we list the field identifications
(barring charge conjugates).
\TABLE[t]{\footnotesize
\renewcommand{\arraystretch}{1.25}
\begin{tabular}{|c|c|c|c|c|c|}
\hline
sector & $(l,m,s)$  &   $h$  & $q$  &    Field   & Comment\\
\hline
NS & $(0,0,0)$   &       0 &  0      &  $V_0 I$               & Identity \\
R & $(1,-2,-1)$ &    3/40 &  +1/10  &  $V_{+1/10}\sigma$     &  RGS \\
NS & $(2,0,0) $  &     2/5 &     0   &  $V_{0}\epsilon$       & \\
R & $(3,-2,-1)$ &   27/40 &   +1/10 &  $V_{+1/10}\rho$       & \\
NS & $(1,+1,0)$  &    1/10 &   +1/5  &  $V_{+1/5} \sigma^*$   & CPO \\
R & $(0,-1,-1)$ &    3/40 &  +3/10  &  $V_{+3/10} I$         & RGS \\
NS & $(3,+1,0)$  &    7/10 &  +1/5   &  $V_{+1/5} \rho^*$     & \\
R & $(2,-1,-1)$ &   19/40 &  +3/10  &  $V_{+3/10} \epsilon$  & \\
NS & $(2,+2,0)$  &     1/5 &   +2/5  &  $V_{+2/5} \sigma$     & CPO \\
R & $(1,0,-1)$  &   11/40 &  +5/10  &  $V_{+5/10} \sigma^*$  & \\
NS & $(3,+3,2)$  &     4/5 &  +2/5   &  $V_{+2/5} \rho$       & \\
R & $(3,0,-1)$  &     7/8 &   +5/10 &  $V_{+5/10} \rho^*$    & \\
NS & $(3,+3,0)$  &    3/10 &  +3/5   &  $V_{+3/5} I$          & CPO \\
R & $(2,+1,-1)$ &   19/40 &  +7/10  &  $V_{+7/10} \sigma$          & \\
NS & $(3,+1,0)$  &    7/10 &  +3/5   &  $V_{+3/5} \epsilon$         & \\
R & $(2,+1,-3)$ &   43/40 &  +7/10  &  $V_{+7/10} \rho$            & \\
NS & $(1,+1,2)$  &     3/5 &  +4/5   &  $V_{+4/5} \sigma^*$         & \\
R & $(3,+2,-1)$ &   27/40 &  +9/10  &  $V_{+9/10} I$               & \\
NS & $(3,+1,-2)$ &     6/5 &  +4/5   &  $V_{+4/5} \rho^*$           & \\
R & $(1,-2,+3)$ &   43/40 &  +9/10  &  $V_{+9/10} \epsilon$        & +9/10 = -11/10(mod2) \\
NS & $(0,0,2)$   &     3/2 &    1    &  $V_{+1}R + V_{-1}\rho^*$    & G (ws susy) \\
NS & $(2,0,2)$   &    9/10 &    1    &  $V_{+1}\sigma + V_{-1}\sigma^*$& \\
\hline
\end{tabular}\caption{The sectors of the $(k=3)^5$ model.}}

\end{document}